\documentclass{article}
\usepackage{graphicx} 
\usepackage{geometry}
\usepackage{amsmath}
\usepackage{braket}
\usepackage{xcolor}
\usepackage{upgreek}
\usepackage{lineno}
\usepackage{authblk}
\usepackage{amssymb}

\usepackage[sorting=none]{biblatex}
\begin{document}

\title{Ultrafast single-photon detection using nanophotonic parametric amplifiers}
\author[1, \dag]{Elina Sendonaris}
\author[2, \dag]{James Williams}
\author[2,3, 4, 5]{Rajveer Nehra}
\author[2]{Robert Gray}
\author[2]{Ryoto Sekine}
\author[2]{Luis Ledezma}
\author[1,2,*]{Alireza Marandi}

\affil[1]{Department of Applied Physics, California Institute of Technology, Pasadena, 91125, California, USA}
\affil[2]{Department of Electrical Engineering, California Institute of Technology, Pasadena, California 91125, USA}
\affil[3]{Department of Electrical and Computer Engineering, University of Massachusetts Amherst, Amherst, Massachusetts 01003, USA}
\affil[4]{Department of Physics, University of Massachusetts Amherst, Amherst, Massachusetts 01003, USA}
\affil[5]{College of Information and Computer Science, University of Massachusetts Amherst, Amherst, Massachusetts 01003, USA}
\affil[$\dag$]{These authors contributed equally.}
\affil[*]{marandi@caltech.edu}

\maketitle

\begin{abstract}
Integrated photonic quantum information processing (QIP) has advanced rapidly due to progress in various nanophotonic platforms. Single photon detectors have been the subject of intense study due to their ubiquity in QIP systems, yet many state-of-the art detectors operate at cryogenic temperatures under vacuum and suffer from long dead times. We propose and demonstrate a single photon detection scheme based on optical parametric amplification in nanophotonic lithium niobate (LN) combined with a classical photodetector. We use quantum detector tomography and experimentally demonstrate an efficiency of 26.5\% with a 2.2\% dark count rate. We show that by improving the nonlinearity-to-loss ratio in nanophotonics and using homodyne detection on a squeezed pump, the detector can achieve 69\% efficiency with 0.9\% dark count rate. The detector operates at room temperature, has no intrinsic dead time, and is readily integrated in LN nanophotonics, in which many other components of photonic QIP are available. Our results represent a step towards all-optical ultrafast photon detection for scalable nanophotonic QIP.
\end{abstract}

\section*{Introduction}
Photonic QIP has been the subject of intense research, due to photons’ long coherence time and their minimal interaction with thermal noise. Particularly, QIP systems such as measurement-based quantum computing \cite{raussendorf_one-way_2001,larsen_deterministic_2021,bartolucci_fusion-based_2021}, quantum computing with linear optics \cite{knill_scheme_2001,menicucci_temporal-mode_2011,menicucci_universal_2006}, and quantum simulations \cite{javid_chip-scale_2023} have been widely studied recently. These systems contain three crucial elements: state generation, state manipulation, and detection. Single photon detectors (SPDs) are the most widely used detectors in photonic QIP, for both determining the outcome of a computation in discrete-variable (DV) QIP \cite{knill_scheme_2001, o2007optical} and for generating non-Gaussian states, necessary for universal continuous-variable (CV) QIP \cite{lloyd1999quantum, baragiola_all-gaussian_2019}, such as Schr\"odinger cat and kitten states \cite{ourjoumtsev2006generating} and GKP states \cite{gottesman2001encoding, konno2024logical}. However, well-performing single photon detection is quite challenging to implement due to photons’ low energies and the prevalence of thermal noise in the environment \cite{hadfield_single-photon_2009}.

Some of the most widely used single photon detectors include superconducting nanowire single photon detectors (SNSPDs) \cite{esmaeil_zadeh_superconducting_2021}, transition-edge sensors (TESs) \cite{lita2008counting}, and single photon avalanche detectors (SPADs) \cite{ceccarelli_recent_2021}. SNSPDs benefit from their high efficiency \cite{reddy_superconducting_2020, marsili2013detecting}, large bandwidths \cite{resta_ghz_2023, grunenfelder_fast_2023}, low dark count rates \cite{shibata_ultimate_2015}, and low timing jitter \cite{korzh_demonstration_2020}. However, they suffer from long dead times and operate at cryogenic temperatures and under vacuum. TESs also operate in cryogenic vacuum conditions and can distinguish between different numbers of photons while having high efficiencies but suffer from a longer reset time compared to SNSPDs \cite{lita2008counting, lau2023superconducting}. SPADs operate at room temperature and pressure but tend to have higher dark count rates and lower efficiencies than SNSPDs \cite{ceccarelli_recent_2021}. For all of these detectors, the wavelength range at which they operate is limited by the materials used, and their performance drops off at wavelengths longer than near-infrared (NIR). 

Moving QIP systems to integrated photonics is of growing interest due to the stability and compactness of the components and the scalability in manufacturing \cite{zhu_integrated_2021,siew_review_2021,lin_advances_2020}. Quantum photonics can benefit from the low loss and strong light-matter interaction offered by integrated platforms \cite{moody2022roadmap, wang2020integrated}. Currently, single photons on photonic chips are detected by either outcoupling them to SPDs off chip \cite{arrazola_quantum_2021} or integrating the detector onto the photonic chip and placing it in a cryostat \cite{sayem_lithium-niobate--insulator_2020}. However, outcoupling photons from a chip leads to a lower total detection efficiency due to the outcoupling loss, and integrating an SNSPD onto a photonic chip creates fabrication and performance challenges.

In this work, we utilize ultrahigh gain nanophotonic OPAs to experimentally realize a single photon OPA detector (OPAD), and analyze its prospects for QIP. Spatial confinement in nanophotonics combined with dispersion engineering enables utilization of ultrashort pulses and provides higher peak intensity, leading to higher nonlinear interaction efficiencies. Recently, nanophotonic optical parametric amplifiers (OPAs) with gains of up to 100 dB/cm have been demonstrated \cite{ledezma_intense_2022}, opening up new avenues for quantum state engineering and measurement \cite{nehra_few-cycle_2022,williams2024ultrashort, kalash_wigner_2023, yanagimoto_mesoscopic_2023}. Optical parametric amplification has been used to both generate and measure measure squeezing over large bandwidths \cite{shaked2018lifting, nehra_few-cycle_2022, kalash_wigner_2023}. Here, we extend the utilization of OPAs into single-photon detection and analyze the OPAD's performance using using the positive operator-valued measure (POVM) framework and quantum detector tomography \cite{feito_measuring_2009}. We experimentally implement such a detector using a periodically poled thin-film lithium niobate (TFLN) waveguide OPA and a fast InGaAs photodetector. We also present an advanced variant of the OPAD and show through simulations that Schr\"odinger's kitten states can be created using such an OPAD for photon subtraction from squeezed vacuum.

\begin{figure}
    \centering
    \includegraphics[width=0.64\textwidth]{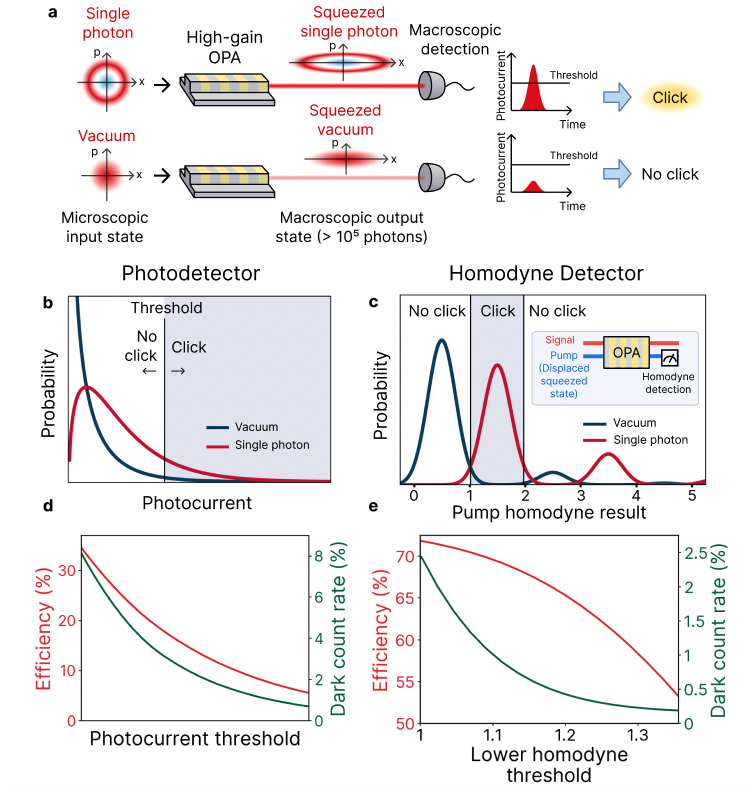}
    \caption{Detecting single photons with an OPA and a macroscopic detector. (a) The concept of the single photon OPAD. A microscopic quantum state is amplified in a waveguide OPA to levels detectable by a macroscopic detector. A threshold is applied to the detected pulse to categorize the signal as single photon or vacuum. (b) Photocurrent distribution from a photodetector when the input to the OPA is vacuum (dark blue) or a single photon (red). An example click threshold is shown. (c) Pump homodyne result distribution of vacuum (dark blue) and a single photon (red) as a function of the homodyne measurement result of the pump's squeezed quadrature. Inset: schematic of the homodyne detection scheme. (d) Efficiency and dark count rate as a function of photocurrent threshold. The specific threshold is dependent on the OPA gain and photodetector, hence is left unlabeled. (e) Efficiency and dark count rate as a function of lower homodyne result threshold, with an upper threshold of 2.}
    \label{fig:intro}
\end{figure}

Figure \ref{fig:intro}(a) shows the concept of the OPAD, which uses an OPA followed by a macroscopic detector. We consider two macroscopic detection methods: a photodetector and a homodyne detector on the pump after the OPA \cite{yanagimoto_quantum_2023}. In both detection schemes, we define a thresholding criteria for a click, as shown in Fig. \ref{fig:intro}(b,c). For a single classical photodetector, this criteria is a single threshold in the photocurrent (Fig. \ref{fig:intro}(c)) and for the homodyne measurement, the criteria is a range of outputs in the pump homodyne. In Fig. \ref{fig:intro}(d,e), we show numerically how the single-photon detection efficiency and dark count rate are affected by the thresholding criteria for these two schemes.

The performance of the photodetector OPAD is limited by how well the system can distinguish between vacuum and a single photon, which is not perfect due to the Gaussian noise erasing the parity information of those states. However, a squeezed single photon has more probability distributed at high photon numbers, and hence photocurrent, than squeezed vacuum does. Therefore, setting the threshold involves a trade off between two sought-after qualities in a single photon detector: low dark counts, which occur when measuring the amplified vacuum results in a photocurrent above the threshold and causes a click; and high photon detection efficiency, which is the probability that measuring a squeezed single photon results in a click. This tradeoff can be seen in Fig. \ref{fig:intro}(d), with both the efficiency and dark count rate falling as the threshold increases.

Using a homodyne detector on the pump after the OPA, akin to the quantum non-demolition (QND) scheme from \cite{yanagimoto_quantum_2023}, shown in the inset of Fig. \ref{fig:intro}(c), can substantially enhance the OPAD's performance. With larger $g/\kappa$ (the ratio of the coupling between the fundamental and second harmonic to the single-photon loss) and fewer modes \cite{yanagimoto_mesoscopic_2023} than exist in current OPAs, one could operate an OPAD in the pump-depletion regime, in which the pump and signal become entangled, and non-Gaussian and Wigner negative features begin to emerge. In this regime, it is possible to project the signal state onto a squeezed Fock state through homodyne conditioning on the depleted pump, which is initially a displaced squeezed state. We simulated this OPAD scheme with the parameters from \cite{yanagimoto_quantum_2023} (normalized interaction time $gt = 1$, pump squeezing of 3 dB, and single-mode OPA squeezing of 3.88 dB), finding a 69.2\% efficiency and 0.9\% dark count rate for certain click categorization limits on the homodyne measurement, corresponding to the shaded region between vertical lines in Fig. \ref{fig:intro}(c). The purity of the state resulting from a click is 0.89. Such an OPAD implements a non-Gaussian detection and could be used to create universal CV QIP resource states. The $g/\kappa$ needed to realize this detection is proportional to the pump squeezing, which in this case corresponds to a requirement of $g/\kappa \gtrsim 1/4$ (see Supplemental Information for more details).

\begin{figure}
    \centering
    \includegraphics[width=0.8\textwidth]{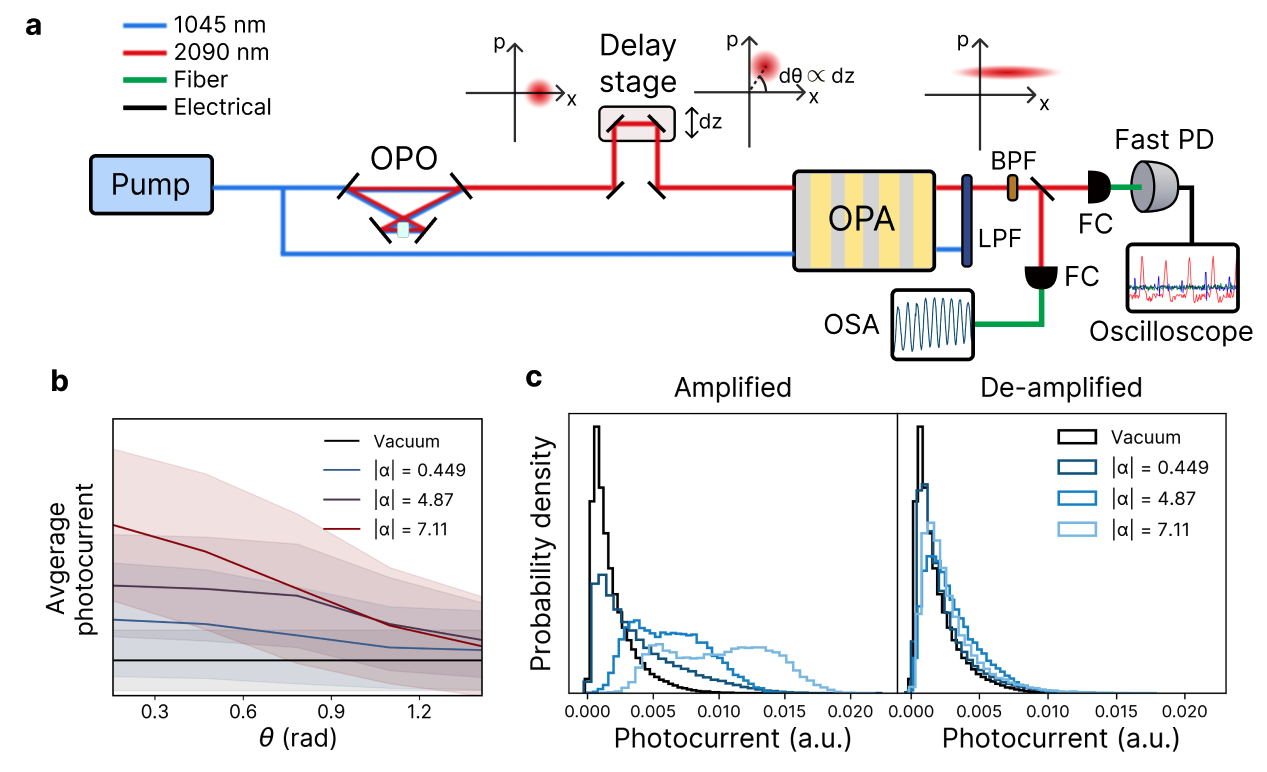}
    \caption{Experimental quantum tomography of an OPA-based detector. (a) Experimental setup for quantum tomography, including Wigner functions of the generated coherent and coherent squeezed states (not to scale). BPF: band-pass filter; LPF: low-pass filter; PD: photodetector; FC: fiber coupler; OSA: optical spectrum analyzer. (b) Average photocurrent of pulses as a function of coherent state angle. Shaded regions indicate the standard deviation of the pulses' photocurrent distribution. (c) Histogram of the pulse photocurrents for amplified and de-amplified coherent states and vacuum.}
    \label{fig:exp_data_ex}
\end{figure}

\section*{Measurement and results}
Using quantum detector tomography \cite{feito_measuring_2009, lundeen_tomography_2009, nehra_characterizing_2020}, we characterized an OPAD made up of an integrated TFLN waveguide OPA (2.4-mm long) and an InGaAs photodetector, as shown in Fig. \ref{fig:exp_data_ex}(a). The waveguide was dispersion engineered for minimal group velocity mismatch between the signal and the pump and for minimal group velocity dispersion at the signal's and pump's frequencies \cite{ledezma_intense_2022}. To perform detector tomography, we sent coherent states at 2.09 $\upmu$m of various amplitudes and phases into the OPA and modulated the phase. The coherent states were generated by a free-space optical parametric oscillator (OPO).

Histograms of the amplified and de-amplified coherent state photocurrents can be seen in Figures \ref{fig:exp_data_ex}(c), respectively. We classified any pulse with a phase $\theta \leq 0.15$ rad as amplified and any pulse with $|\pi/2 - \theta| \leq 0.15$ rad as de-amplified, as we found this to be a reasonable bin size through variation of the total number of bins. The amplified pulses have a higher average photocurrent and variance than amplified vacuum, and the de-amplified pulses have a similar distribution to amplified vacuum, as expected. The cause of the double-bumped distribution of $|\alpha| = 2.13$ in Figure \ref{fig:exp_data_ex}(c) is likely pump depletion, which leads to gain saturation and a bunching of the tail of higher measurements as we see. The average photocurrent as a function of the coherent state angle $\theta$, with the data binned into 5 angle bins between 0 and $\pi/2$, is plotted in Figure \ref{fig:exp_data_ex}(b). The shaded regions indicate the standard deviation of the data. As $\theta$ approaches $\pi/2$, the average and standard deviation approaches that of a squeezed vacuum state, corresponding to the de-amplification of the coherent states.

\begin{figure}
    \centering
    \includegraphics[width=\textwidth]{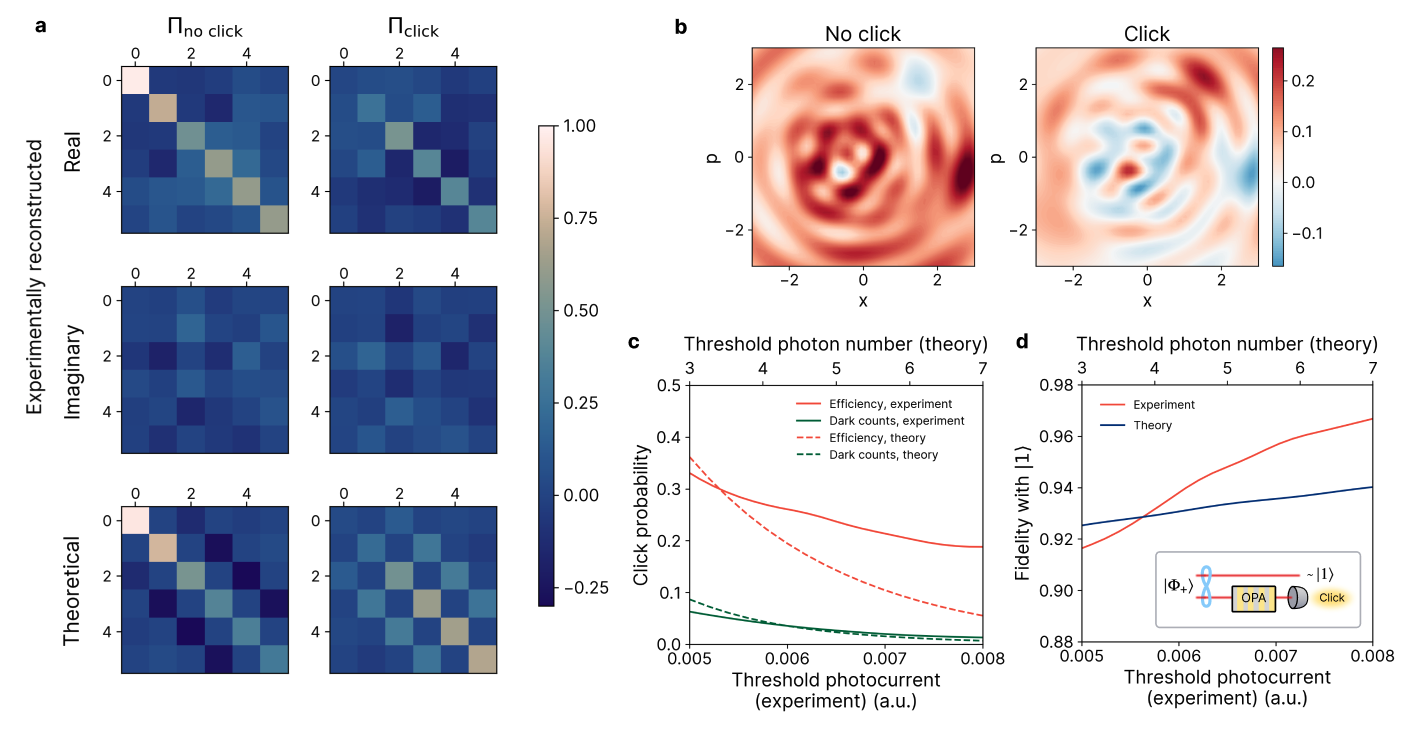}
    \caption{Detector tomography results and analysis.(a) The POVM elements of the OPAD, reconstructed through detector tomography (first and second rows) and theoretically calculated using QuTiP (third row). (b) Experimentally reconstructed Wigner functions of the no click and click POVM elements. (c) The efficiency and dark count rate of the detector for both experimentally reconstructed and theoretically calculated POVMs as a function of the threshold photocurrent or photon number. (d) The fidelity with a single photon state after one mode of the Bell state $(\ket{00} + \ket{11})/\sqrt{2}$ is measured with an OPAD vs. the threshold photocurrent or photon number, for both experimental and theoretical POVMs. Inset: diagram of the measurement scheme.}
    \label{fig:povm_quad}
\end{figure}

To obtain the POVM of our OPAD, we used a convex optimization algorithm, similar to the one described in \cite{feito_measuring_2009} (see Methods). The experimentally retrieved and theoretical POVMs are shown in Figure \ref{fig:povm_quad}
(a). The experimentally reconstructed POVM has a dark count probability of 2.2\% and an efficiency of 26.2\%, while the theoretical POVM has a dark count probability of 4.4\% and an efficiency of 22.7\%. The Wigner functions of the experimentally reconstructed POVM elements are shown in Fig. \ref{fig:povm_quad}(b), with the click having negativity in the center. 

The theoretical POVM was calculated for an OPA with a scaled down amount of gain (10 dB) in order to be computationally viable, and with proportional noise. Since more squeezing does not change the POVM qualitatively, it is possible to recover the same POVMs from different amounts of gain by choosing appropriate thresholds (see Supplementary Information for details). We calculated the fidelity $F_n = \text{Tr}((\Pi_{n, \text{theo}}^{1/2} \Pi_{n, \text{exp}} \Pi_{n, \text{theo}}^{1/2})^{1/2})^2$ \cite{feito_measuring_2009}, with the trace of each of the POVM elements normalized to 1, for every combination of $N_{th}$ in the theoretical POVM and photocurrent threshold in the experimental reconstruction. The average fidelity of both elements, $(F_\text{click} + F_\text{no click})/2$, was over 0.72 for certain thresholds. 

To quantify the trade-off between efficiency and dark counts, we swept the cutoff photocurrent in our convex optimization algorithm and swept $N_{th}$ in the theoretical POVM calculation. The efficiencies and dark counts vs. the threshold photon number (for theory) and threshold photocurrent (for experiment) both for the experimental and theoretical POVMs are plotted in Figure \ref{fig:povm_quad}(c), with the threshold photocurrent adjusted to match the threshold photon number proportionally to the gains of the OPAs. The experimental POVM outperforming the theoretical POVM at higher thresholds is an indication that the pump depletion, which causes the bump in the amplified photocurrent distribution, is improving the performance. This is because as the threshold photocurrent increases, especially above 0.006, the detection probability does not decrease as quickly as for a squeezed coherent state with no pump depletion.

We considered an example of DV detection in which one mode of a Bell state $\ket{\Phi_+} = (\ket{00} + \ket{11})/\sqrt{2}$ is sent to the OPAD. We calculated the fidelity of the other mode with a single photon ($\ket{1}$) as a function of the detection efficiency using the theoretical and experimentally reconstructed POVMS. The results are plotted in Figure \ref{fig:povm_quad}(d). As expected, the theoretical curve indicates a decrease in fidelity as the detection efficiency increases. This is due to the higher dark count probability. The experimental POVM has a higher fidelity than the theoretical fidelity, which can be explained by the pump depletion.

\section*{Discussion}
\begin{figure}
    \centering
    \includegraphics[width=0.5\textwidth]{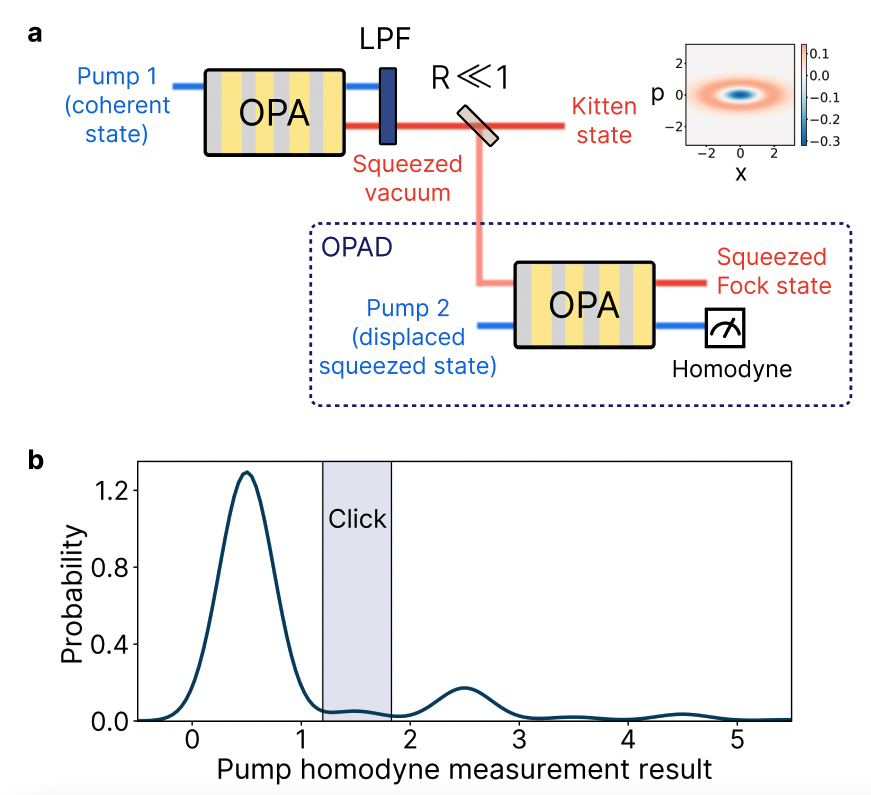}
    \caption{Using a pump homodyne OPAD to create a kitten state. (a) Diagram of the scheme. The first OPA creates squeezed vacuum and the second OPA is followed by a homodyne measurement on the depleted pump, triggering a click. The Wigner function of the kitten state is shown on the upper right. (b) The detection probability as a function of the pump homodyne measurement result, with the detection window considered to create the kitten state shaded.}
    \label{fig:qnd_det}
\end{figure}

The main benefits of the OPAD are its speed, room-temperature operation, integrability, and flexibility in wavelength. The speed of detection is limited only by the laser repetition rate, the photodetection system's electronic bandwidth, and the width of the pulses. We detected pulses with an 18GHz detector, which is on the high end of the resolution of SNSPDs and SPADs \cite{chen_photon-trapping-enhanced_2023, resta_ghz_2023}. We operated at a repetition rate of 250 MHz, but there is no fundamental physical obstacle to faster operation. Typically, the efficiency of SNSPDs and SPADs decreases as the wavelength increases due to bandgap-related limitations \cite{dello_russo_advances_2022}, however, the OPAD can be designed to operate flexibly within the transparency window of LN spanning from the visible to the mid-IR. Furthermore, the simplicity of the design means that this detector can be readily integrated into the fast-evolving nonlinear nanophotonic circuits in LN. Additionally, our OPAD scheme is tolerant to losses after the OPA, as the loss affects all amplified states equally and can thus be accounted for by setting a lower threshold photocurrent.

The proposed OPAD can be used as an effective photon subtraction scheme for generation of non-Gaussian states with further improvements on the gain and material loss on nanophotonic platforms. We consider creating a Schr\"odinger kitten state through photon subtraction \cite{ourjoumtsev2006generating} a pump homodyne OPAD similar to the QND scheme and parameters in \cite{yanagimoto_quantum_2023}, as shown in Fig. \ref{fig:qnd_det}(a) (see Methods for details). In this scheme, when the pump homodyne measurement is in a specific range, a click is triggered, projecting the initial squeezed vacuum onto a kitten state. The ensemble average output state purity is numerically estimated to be 0.98 (shown in Fig. \ref{fig:qnd_det}(b)),  and the fidelity between the squeezed photon subtracted two-mode squeezed state and a photon subtracted squeezed state is 0.96, with a click probability of 2.4\%. Experimental realization of such kitten states requires further advancement in the losses in nanophotonic waveguides and the multimode effects that occur with broadband pulsed operation and dispersion engineering \cite{wasilewski_pulsed_2006}. The propagation loss limits the $g/\kappa$ of current realizations of OPAs to much less than one. Significant non-Gaussianity can be observed only when $g/\kappa$ is around the order of unity (see Supplemental Information).

\section*{Conclusion}
We have proposed and demonstrated an OPAD on a TFLN nanophotonic chip.  Our detector works at room temperature and pressure, and as the optical parametric amplification happens instantaneously, the speed of the detector is limited by how low the dispersion is, which determines the minimum duration of the pulses. Through tailoring of the material, periodic poling, and dispersion engineering, it is possible to detect photons from visible wavelengths into the MIR. As an example of the OPAD's utility for DV QIP, we simulated a measurement using our experimental POVM on one mode of a Bell state and showed that for a detection efficiency of 30\%, the fidelity of the other mode with a single photon is over 92\%, showing how this system is useful for ultrafast DV state manipulations. We also showed numerically that it is possible to create a Schr\"odinger kitten state with a fidelity of 0.96 with an OPAD that uses homodyne detection on the pump. Our results suggest that single photon all-optical detection using second-order optical nonlinearity is a promising direction for ultrafast integrated photonic quantum information processing.

\section*{Methods}
\subsection*{Experimental setup}

We send coherent states of known amplitude and phase created by an OPO into the OPAD to gather data for POVM reconsruction. We pump both the OPO and the OPAD with 75-fs pulses at 1.045 $\upmu$m at 250 MHz from a mode-locked laser. The coherent states' amplitudes were adjusted with a neutral density filter. The phase of the coherent state was swept using a piezoelectric adjustable delay line. To reduce the multi-mode effects in the amplified light and filter the pump out, we used a 48-nm bandwidth band-pass filter centered at 2.09 $\upmu$m. In addition to this filter, we used one 1330-nm long-pass filter and two 1500 nm long-pass filters, which provided 150 dB of total pump rejection. Both the coherent states and the pump pulses were coupled onto the chip using a reflective objective and outcoupled onto a tapered single-mode fiber.

Using a 90/10 beam-splitter, we recorded both the individual pulse photocurrent amplitudes and a slower signal from an optical spectrum analyzer (OSA) containing the amplification and de-amplification envelope of the signal (amp/de-amp) as the coherent state's phase was swept. The pulse data was recorded using an InGaAs photodetector with an 18 GHz bandwidth with an 80Gs/s 40GHz oscilloscope, enough to resolve individual electronic pulses. The amp/de-amp envelope was recorded on a seperate slow photodetector. This curve was used to determine the phase of the coherent state relative to the pump. We recorded data from over 4 million pulses and collected over 140,000 pulses of optical parametric generation data, in which the signal is the vacuum state.

In order to characterize the amplitude of the coherent states on-chip, we sent the pulses into an SNSPD optimized for operation at 2 $\upmu$m. Knowing the chip output loss from OPG measurements (see Supplemental Information) and the loss from the filters allowed us to calculate the amplitude from the fraction of pulses that caused a click on the SNSPD.

\subsection*{Quantum detector tomography}
We calculated the theoretical click and no-click POVM matrices for a squeezer with 10 dB of squeezing. We used $\eta = \eta_{pd}\eta_f = 0.314$, for photodetector efficiency $\eta_{pd} = 0.629$ and filter loss $\eta_f = 0.5$. In order to determine the appropriate standard deviation $\sigma$ for 10 dB of squeezing we used the ratio of the standard deviation of the electronic noise to the average photocurrent for squeezed vacuum as a metric for the amount of noise. This metric was 0.172 for our setup and the average photon number for a 10 dB squeezed vacuum contains 2.02 photons, so for the theoretical POVM we used $\sigma = 2.02 \times 0.172 = 0.347$ photons. We summed over photon numbers up to 150 since at 10 dB the average photon number of squeezed vacuum is 2.02 and the probability amplitude at 150 photons is $10^{-7}$.

To experimentally reconstruct the POVM of our OPAD, we used the convex optimization process described in \cite{feito_measuring_2009}, but with the full POVM matrices rather than just the diagonal entries, since this is a phase-sensitive measurement. Because of this, we must solve a tensor equation rather than just a matrix one. We seek to invert the following equation:
\begin{align}
    P_{i, n} = F_{i, k, p} \Pi_{k, p, n}, \label{eqn:povm}
\end{align}
where $P_{i, n} = \bra{\alpha_i}\pi_n\ket{\alpha_i}$ is the experimentally determined probability of getting a click or not as a function of the coherent state amplitude $\alpha_i$, and $F_{i, k, p} = e^{-|\alpha_i|^2}(\alpha_i^*)^k \alpha_i^p/\sqrt{k!p!}$ are the prefactors associated with the type of state being sent in (in our case, coherent states). $\Pi_{k, p, n} = \sum_{k, p} \theta^{(n)}\ket{k}\bra{p}$ are the POVM  matrices we are trying to find by performing a convex optimization which minimizes $||P - F\Pi||_2$. The indices $k$ and $p$ are the indices of the matrices, $n$ refers to which measurement result ($n = 0$ for no click, $n = 1$ for a click), and $i$ indexes the coherent state amplitude.

The optimization also involves a regularization constraint which smooths the diagonal entries of the POVM, which takes the form of 
\begin{align}
    R = \sum_{k, n} (\theta^{(n)}_{k + 1, k + 1} - \theta^{(n)}_{k, k})^2.
\end{align}
We can then perform the minimization over
\begin{align}
    \min(||P - F\Pi||_2 + \gamma R),
\end{align}
for some regularization constant $\gamma$. Here we used $\gamma = 10^{-4}$, a relatively low value, in order to prevent over-smoothing of the POVM, as we expect some sharp characteristics from $\ket{0}\bra{0}$ to $\ket{1}\bra{1}$ indicating the click-like character of the detector (for more details see the Supplemental Information).

We used data from coherent states with amplitudes $|\alpha|$ = 0.449, 4.87, and 7.11 and binned the phases into five bins between 0 and $\pi/2$, as shown in Fig. \ref{fig:exp_data_ex}(c), in order to discretize the data enough for the convex optimization. We found that varying the bin sizes did not create large variations in the output, so we concluded that five bins was enough. We used CVXPY, a Python library for solving convex optimization problems, to invert Eqn. \ref{eqn:povm} \cite{cvxpy}. 

\subsection*{Homodyne OPAD}
We consider an OPAD with the homodyne measurement on the pump outlined in \cite{yanagimoto_quantum_2023}, which projects the signal onto a mixture of squeezed Fock states based on a homodyne measurement of a depleted pump. The parameters used in our simulations are the same as those considered in that paper. For an OPA Hamiltonian in a displaced frame of $\hat{H} = g(\hat{a}^{\dag 2}\hat{b} + \hat{a}^2\hat{b}^\dag) + \delta\hat{a}^\dag\hat{a} + \frac{r}{2}(\hat{a}^{\dag 2} + \hat{a}^2)$, as for Figure 1 of that paper, we use $\sqrt{1 - r^2/\delta^2} = 150$ and $\tilde{g}/g = \sinh(\tanh^{-1}(r/\delta)) = 1$, which leads to the Fock states post-measurement having 3.88 dB of squeezing. We also set the total interaction time $gt = 1$. For the purity calculations, we consider a pump with 3 dB squeezing.

When we consider creating a Schr\"odinger kitten state using squeezed photon subtraction from squeezed vacuum using this scheme, we simulate a 5 dB squeezed vacuum going through a beam-splitter with reflectivity 0.2, with the OPAD on the reflected port. For our parameters, the measurement is effectively a squeezed Fock state subtraction, with the number of photons dependent on the pump homodyne detection result. We set the limits of the detected range of pump homodyne measurements which trigger a click such that the ensemble average output state purity is 0.98 and the average fidelity between this squeezed-photon-subtracted two-mode squeezed state and a photon-subtracted squeezed state is over 0.96. The probability of getting a detection result in that range is 2.4\%. 

\section*{Data availability}
The data used to generate the plots and results in this paper are available from the corresponding author upon reasonable request.

\section*{Code availability}
The code used to analyze the data and generate the plots for this paper is available from the corresponding author upon reasonable request.

\printbibliography

\section*{Acknowledgements}
Device nanofabrication was performed at the Kavli Nanoscience Institute (KNI) at Caltech. The authors gratefully acknowledge support from ARO grant no. W911NF-23-1-0048, NSF grant no. 1846273, 1918549, 2139433, AFOSR award FA9550-23-1-0755, the center for sensing to intelligence at Caltech, the Alfred P. Sloan Foundation, and NASA/JPL. The authors wish to thank NTT Research for their financial support. RN would like to thank Dr. Saman Jahani for fruitful discussions. ES would like to thank Edwin Ng and Ryotatsu Yanagimoto for advice on g/$\kappa$ calculations.

\section*{Author information}
These authors contributed equally: Elina Sendonaris and James Williams

\subsection*{Contributions}
E.S., J.W., R.N., and A.M. conceived the idea and designed the experiments. E.S. and J.W. carried out
the experiments with help from R.G.. E.S. carried out the numerical simulations with help from L.L.,
J.W. and R.N. R.S. fabricated the device. E.S. wrote the manuscript with input from all authors. A.M.
supervised the project.

\section*{Ethics declarations}
\subsection*{Competing interests}
A.M., R. N., L.L., and R.S. are inventors on the U.S. patent application (application number 18/155,444). R.S., L.L. and A.M. are involved in developing photonic integrated nonlinear circuits at PINC Technologies Inc. R.S., L.L. and A.M. have an equity interest in PINC Technologies Inc.

\appendix
\section{OPAD POVM}
The POVM of an OPA followed by a photodetector with imperfect efficiency and Gaussian electronic noise, given by \cite{davis_conditional_2021}, is 
\begin{equation}
    \Pi_n = \sum_{m \geq 0} P(n|m) \hat{S}^\dag(\xi) \ket{m}\bra{m} \hat{S}(\xi)
    \label{eqn:pi_n}
\end{equation}
where $P(n|m)$ is the probability that the photodetector will register $n$ photons when $m$ photons arrive at the detector. We assume that the OPA is operating at degeneracy in the low pump-depletion, linear gain regime, so that it implements the single-mode squeezing operator $\hat{S}(\xi)$, for squeezing parameter $\xi$. Because the squeezing operator is phase-sensitive, this is a phase-sensitive detector. $P(n|m)$ depends on the photodetector's efficiency $\eta$ and sources of Gaussian noise as 
\begin{equation}
    P(n|m) = \sum_{q \geq 0} P_\sigma(n|q)P_\eta(q|m),
    \label{eqn:p_nm}
\end{equation}
where
\begin{equation}
    P_\sigma(n|q) = \exp \left( \frac{-(n - q)^2}{2\sigma^2} \right)
    \label{eqn:p_noise}
\end{equation}
is the probability of detecting $n$ photons given $q$ photons arrived at the detector and there is Gaussian noise with standard deviation $\sigma$, representing the electronic noise. The term
\begin{equation}
    P_\eta(q|m) = \frac{m!}{q!(m - q)!} \eta^q (1-\eta)^{m - q}
    \label{eqn:p_loss}
\end{equation}
is the probability that $q$ photons are detected after $m$ photons pass through a beam-splitter with transmission $\eta$, which models the finite efficiency $\eta \leq 1$ of the photodetector.

We can model a click detector by setting a threshold number of photons $N_{th}$ above which the detector registers a click. Then, the click POVM is
\begin{equation}
    \Pi_\text{click} = \sum_{n \geq N_{th}} \Pi_n
    \label{eqn:pi_click}
\end{equation}
and $\Pi_\text{no click} = \mathbf{1} - \Pi_\text{click}$. The probability of getting result $k$ is $p_k = \text{Tr}(\rho \Pi_k)$, so the single photon efficiency is the (1,1) matrix element of $\Pi_\text{click}$ in the Fock basis, and the dark count rate is the (0,0) element of the same matrix. All the summations are all over all photon numbers, but for computational purposes it suffices to sum over photon numbers within the range where there is significant probability density based on the squeezing parameter and noise standard deviation.

\section{Chip dimensions and characterization}
We used a thin-film lithium niobate (TFLN) chip with a thickness of 700 nm, waveguide etch depth of 345 nm and width of 1.85 $\mu$m. The poling period is 5.22 $\mu$m. We estimate that the group velocity mismatch between the fundamental and second harmonic is 5.2 fs/mm, and the group velocity dispersion at the fundamental is 120.1 fs$^2$/mm and at the second harmonic it is 28.4 fs$^2$/mm.

\begin{figure}
    \centering
    \includegraphics[width=0.7\textwidth]{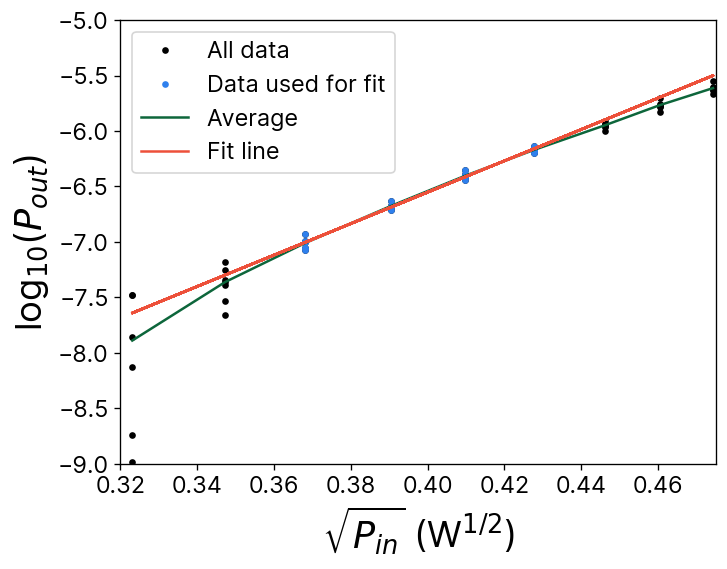}
    \caption{OPG power as a function of pump power. We used a subset of the data which is in the undepleted pump regime and also with high enough power to consistently register on our detector, shown in blue. The average power is shown as the green line for illustration purposes. The fit equation with the parameters we found is shown in red.}
    \label{fig:opg_curve}
\end{figure}

We characterized the chip's gain and output coupling by measuring the optical parametric generation (OPG) power as a function of pump power. We did so by blocking the signal path from the OPO and adjusting the pump power using a half-wave plate and polarizing beam-splitter. The data we collected is shown in Fig. \ref{fig:opg_curve}. The equation we fit to is
\begin{align}
    P_{OPG, off} &= \hbar \omega_s \eta_{OC} f_{rep}\sinh^2(L\sqrt{\eta_{NL} \eta_{IC}P_{pump, off}}) \label{eqn:opg_curve} \\
    &\approx \frac{\hbar \omega_s \eta_{OC} f_{rep}}{4} e^{2L\sqrt{\eta_{NL}\eta_{IC}P_{pump, off}}},
\end{align}
where the last approximation is in the large-gain regime. $P_{OPG, off}$ is the OPG power off-chip, $P_{pump, off}$ is the pump power off-chip, $\eta_{OC}$ is the output coupling efficiency of the chip, $\eta_{IC}$ is the input coupling efficiency, $\omega_s$ is the signal frequency, $f_{rep}$ is the laser repetition rate, $L$ is the length of the OPA, and $\eta_{NL}$ is the normalized second harmonic efficiency. We fit to a linearized version of the equation, $\ln(P_{OPG, off}) = \ln(\hbar \omega_s \eta_{OC} f_{rep}/4) + 2b\sqrt{P_{pump, off}}$. We find $\eta_{OC} = 0.102$ and $b = L\sqrt{\eta_{NL}\eta_{IC}} = 16.3$ with standard deviations of the fit $\sigma_{\eta_{OC}} = 0.027$ and $\sigma_b = 0.328$. Thus, the output coupling loss is 9.9 dB. The transmission loss through the chip is 25 dB, which leads to an input coupling loss of approximately 15.1 dB. This is likely due to the mode mismatch between the output of the reflective objective and the waveguide. We find $\eta_{NL} = (1.46 \pm 0.06) \times 10^5~\text{W}^{-1}\text{cm}^{-2}$.

\subsection{Extracting g/$\kappa$ from OPG measurements}
The interaction Hamiltonian for 3-wave mixing is
\begin{equation}
    \frac{\hat{H}}{\hbar} = g(\hat{a}^{\dag 2}\hat{b} + \hat{a}^2\hat{b}^\dag)
\end{equation}
for interaction strength $g$, signal annihilation operator $\hat{a}$, and pump annihilation operator $\hat{b}$. In the Heisenberg picture under the undepleted-pump assumption with pump strength $\hat{b} \rightarrow \beta$, for an initial state of vacuum, the signal number operator evolves as $\langle\hat{a}^\dag\hat{a}\rangle(t) = \langle\hat{N}_a\rangle(t) = \sinh^2(2g|\beta|t)$. To compare this to the OPG equation (Eqn. \ref{eqn:opg_curve}), we consider how to convert these quantities into the power measurements. $\langle\hat{N}_a\rangle(t) = P_{OPG, on}/(\hbar\omega_s f_{rep})$ and $|\beta| = \sqrt{\langle\hat{N}_b\rangle} = \sqrt{P_{pump, on}/(\hbar\omega_p f_{rep})}$ where the "on" subscript refers to the on-chip power. Substituting, we have
\begin{align}
    P_{OPG, off} \approx \frac{\hbar\omega_s f_{rep} \eta_{OC}}{4} e^{4gt\sqrt{\eta_{IC}P_{pump, off}/(\hbar\omega_p f_{rep})}}
\end{align}

Comparing to Eqn. \ref{eqn:opg_curve}, we can see $4gt\sqrt{\eta_{IC}P_{pump, off}/(\hbar\omega_p f_{rep})} = 2L\sqrt{\eta_{NL}\eta_{IC}P_{pump, off}}$, so we have $g = L\sqrt{\eta_{NL}\hbar\omega_p f_{rep}}/(2t)$, and with $t = Ln / c$, finally,
\begin{equation}
    g = \frac{c}{2n}\sqrt{\eta_{NL}\hbar\omega_p f_{rep}}.
\end{equation}
From the value of $\eta_{NL}$ from the fit, we have $g = 29$ MHz.

However, we must account for multimode effects in our OPA. The simplest, most naive way of dealing with it is to assume that each mode receives an equal amount of the gain. We calculate the purity $\rho$ from the normalized joint spectral intensity by performing a singular value decomposition and adding the squares of the eigenvalues, and obtain the effective mode number $m_{eff} = 1/\rho \approx 12.6$. We can then calculate $g_{eff} = g / m_{eff} = 2.3$ MHz.

We compare this number to the theoretical value using this equation from \cite{yanagimoto2022temporal}, modified to include the naive multimode effects:
\begin{equation}
    g_{eff} = \frac{4 d_{eff}}{m_{eff}\lambda^3}\sqrt{\frac{2\pi^3\hbar c^3}{n^3 \varepsilon_0 \tilde{V}_{sh}}}
\end{equation}
where $d_{eff}$ is the effective quadratic nonlinear coefficient, which is $d_{33}/(2\pi)$, with $d_{33} = 20$ pm/V and $\tilde{V}_{sh}$ is the effective mode volume divided by $(\lambda/n)^3$, with perfect mode overlap resulting in it being exactly the normalized mode volume. We use the size of the waveguide 400 nm by 1780 nm times the pulse width of $\tau_p$ = 100 fs equating to $\tau_p c/n$ = 1.62 $\mu$m for $n = 1.85$. Assuming perfect overlap and plugging in the parameters for our waveguide and pulse width to get the volume, we get an estimate of $g = 9.0$ MHz. This value is of the same order to the experimental extraction, and the discrepancy may be attributed to the imperfect mode overlap between the first harmonic and second harmonic.

We estimate the loss in the waveguide of 0.23 dB/cm and a group velocity of $v_g = 1.33 \times 10^8$ m/s, which leads to a $\kappa = \alpha v_g$ \cite{yanagimoto2022temporal} of 702 MHz. Thus, $g_{eff}/\kappa \approx 0.0033$.

\begin{figure}
    \centering
    \includegraphics[width=0.85\linewidth]{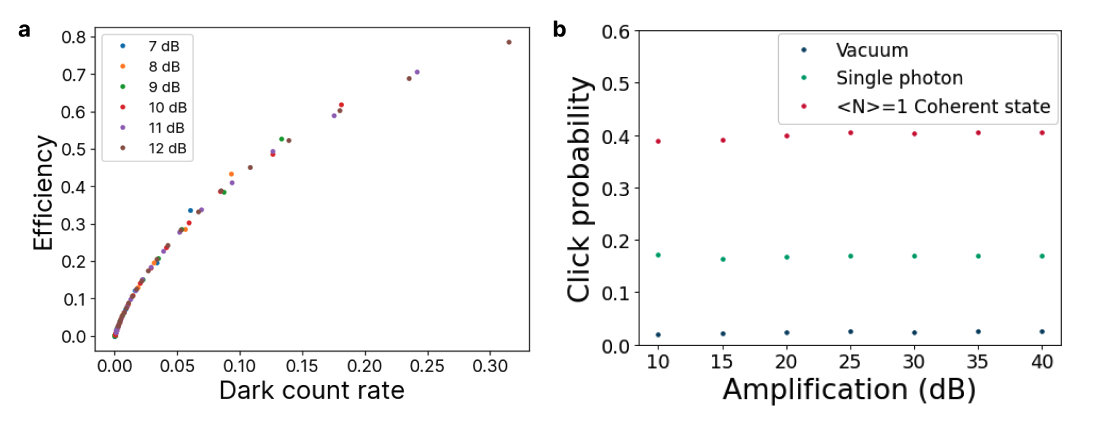}
    \caption{The performance of OPAs with different levels of gain. (a) The efficiency vs. dark count rate of OPADs with OPAs with various levels of gain with Gaussian noise proportional to the gain. The points were generated by sweeping the threshold photon number. (b) Click probability vs. OPA gain for vacuum, single photon, and an $\langle N\rangle = 1$ coherent state, for vacuum click probability (dark count rate) of 2.5\%. Note the constant click probabilities over a wide amplification range.}
    \label{fig:gains}
\end{figure}
\section{High-gain OPA simulations}
The OPA in our theoretical POVM calculations has 10 dB of squeezing. This is clearly below the gain that our actual on-chip OPA has, which is in the tens of dBs. However, we do not have to simulate the actual gain of the OPA because the nature of the POVM does not qualitatively change with higher gain. Squeezing stretches out the photon number distributions but does not change their characteristics beyond 5 dB of squeezing. 

To test whether the gain affected the efficiency-dark count tradeoff, we simulated OPADs with gains between 7 and 12 dB, adjusted the Gaussian noise's standard deviation to be proportional to the average photon number of vacuum squeezed by that parameter, and swept the threshold photon number. The gain does not affect the efficiency-dark count rate curve, as can be seen in Fig. \ref{fig:gains}(a). The difference between the different gains is that for a the same threshold, the OPAD will be on different places on this curve. However, by adjusting the threshold it is possible to achieve the same performance as any other OPAD with different gain.

As another example, we calculated the click probability for input states of a single photon and a coherent state with $\langle N \rangle = |\alpha|^2 =1$ for a threshold which yields a 2.5\% dark count rate (a.k.a. vacuum click probability). As can be seen in Fig. \ref{fig:gains}(b), the rates stay constant over squeezing from 10 to 45 dB, indicating that an OPAD's performance is not based on the OPA's gain, and that as long as the classical detector is sensitive enough to detect the pulses, an appropriate threshold can be set to achieve any performance along the dark count rate-efficiency curve.

Thus, we conclude that it is possible to gain insight from a simulation of an OPAD with 10 dB of gain despite it being much less than the actual gain.

\begin{figure}
    \centering
    \includegraphics[width=0.5\linewidth]{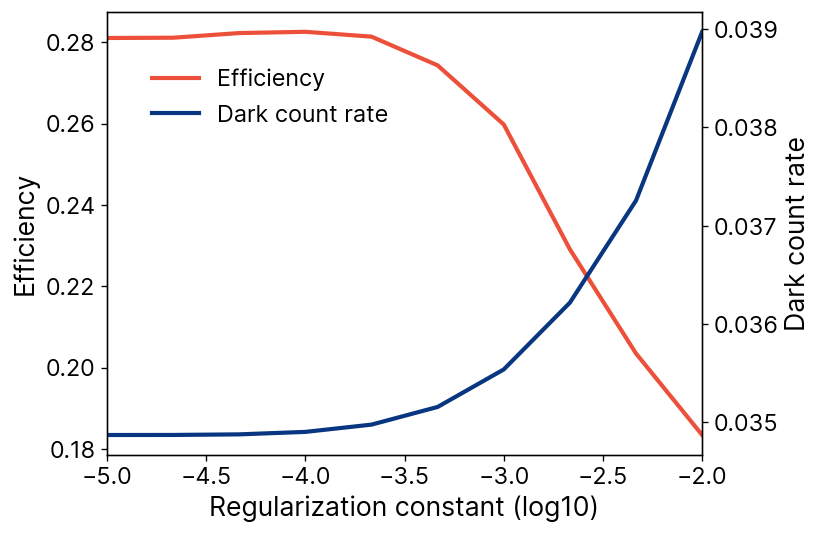}
    \caption{A plot of dark count rate and efficiency as a function of the regularization constant $\gamma$.}
    \label{fig:reg}
\end{figure}

\section{Detector tomography regularization constant}
To invert the matrix equation to extract the POVM, we perform convex optimization to minimize this equation:
\begin{align}
    \min(||P - F\Pi||_2 + \gamma R),
\end{align}
with $\gamma R$ being the regularization to smooth the potentially spiky nature of the diagonal entries of the POVM, with $R  = \sum_{k, n} (\theta^{(n)}_{k + 1, k + 1} - \theta^{(n)}_{k, k})^2$. To ensure we use a value of $\gamma$ not too high such that it starts to introduce artifacts, we swept $\gamma$ and found the dark count rate and efficiency as a function of $\gamma$ (Fig. \ref{fig:reg}). We saw that $10^{-4}$ was within the region such that changing it did not change the result significantly, so we chose that value.

\end{document}